\documentclass{nature}

\usepackage{graphicx}
\usepackage{amsmath}
\usepackage{epsfig}
\usepackage{subfigure}
\usepackage{color}

\DeclareGraphicsExtensions{.pdf}

\title{A Fowler-Nordheim Integrator Can Track the Density of Prime Numbers}

\author{Liang Zhou$^1$, SriHarsha Kondapalli$^2$ \& Shantanu Chakrabartty$^2$}

\begin{document}

\maketitle

\begin{affiliations}
 \item Department of Computer Science and Engineering, Washington University in St. Louis
 \item Department of Electrical and Systems Engineering, Washington University in St. Louis
\end{affiliations}

\begin{abstract}

{\it \textbf{Does there exist a naturally occurring counting device that might elucidate the hidden structure of prime numbers?}} is a question that has fascinated 
computer scientists and mathematical physicists for decades\cite{van2004quantum,latorre2014there}. 
While most recent research in this area have explored the role of the Riemann zeta-function\cite{riemann1859number} in different formulations of statistical mechanics, condensed matter physics\cite{dyson2009birds, planat2011riemann,schumayer2011colloquium} and quantum chaotic systems\cite{bunimovich2005open, schaden2006sign}, the resulting devices (quantum or classical) have only existed in theory or the fabrication of the device has been found to be not scalable to large prime numbers. Here we report for the first time that any hypothetical prime number generator, to our knowledge, has to be a special case of a dynamical system that is governed by the physics of Fowler-Nordheim (FN) quantum-tunneling. 
In this paper we report how such a dynamical system can be implemented using a counting process that naturally arises from sequential FN tunneling and integration of electrons on a floating-gate (FG) device. 
The self-compensating physics of the FG device makes the operation reliable and repeatable even when tunneling-currents approach levels below $10^{-18}$A. 
We report measured results from different variants of fabricated prototypes, each of which shows an excellent match with the asymptotic prime number statistics. 
We also report similarities between the spectral signatures produced by the FN device and the spectral statistics of a hypothetical prime number sequence generator.
We believe that the proposed floating-gate device could have future implications in understanding the process that generates prime numbers with applications in security and authentication\cite{rivest1978method}.
\end{abstract}

Understanding the hidden structure of prime numbers has been a generation-old problem that has spawned multiple unanswered hypotheses, including the celebrated Riemann Hypothesis\cite{dickson2005history}. 
This is because while prime numbers appear to be randomly distributed, they also exhibit some statistical structure, as evident by recent results in this area~\cite{maynard2016chains, stanfordPNAS}. 
The daunting and yet universal nature of this problem has led researchers to ask the question whether there exists a naturally occurring analog or quantum device that might exhibit properties similar to that of prime numbers. 
After all, analog devices can naturally solve transcendental equations\cite{daniel2013synthetic} and statistical mechanics could be used to model NP-hard problems\cite{bounds1987new}. 
For example,  a remarkable analog device was reported in \cite{van1947electro} and was used for estimating the locations of the non-trivial zeros of the Riemann zeta-function. 
However, like many analog implementations, the precision of the fabricated device limited the scalability of the approach towards estimating higher-order zeros. 
Recent research has focused on exploring formulations in statistical, condensed matter and quantum physics\cite{schumayer2011colloquium} where Riemann zeta-function could be naturally embedded. 
Intriguing statistical similarities between the distribution of energy levels of heavy nuclei\cite{firk2009nuclei}, the distribution of eigenvalues of certain random matrices\cite{connes1999trace}, the eigenstates of Hamiltonian operators\cite{sekatskii2007hamiltonian, bender2017hamiltonian, schumayer2008quantum} and the distribution of the location of non-trivial zeros of the Riemann zeta-function have been reported. 
However, many of these mathematical constructs have been theoretical and it is not clear how they could be translated towards synthesizing a practical and an operational device. 

\begin{figure*}[ht]
	\centering
	\begin{minipage}[t]{1\textwidth}
		\centering
		\includegraphics[width=1\textwidth]{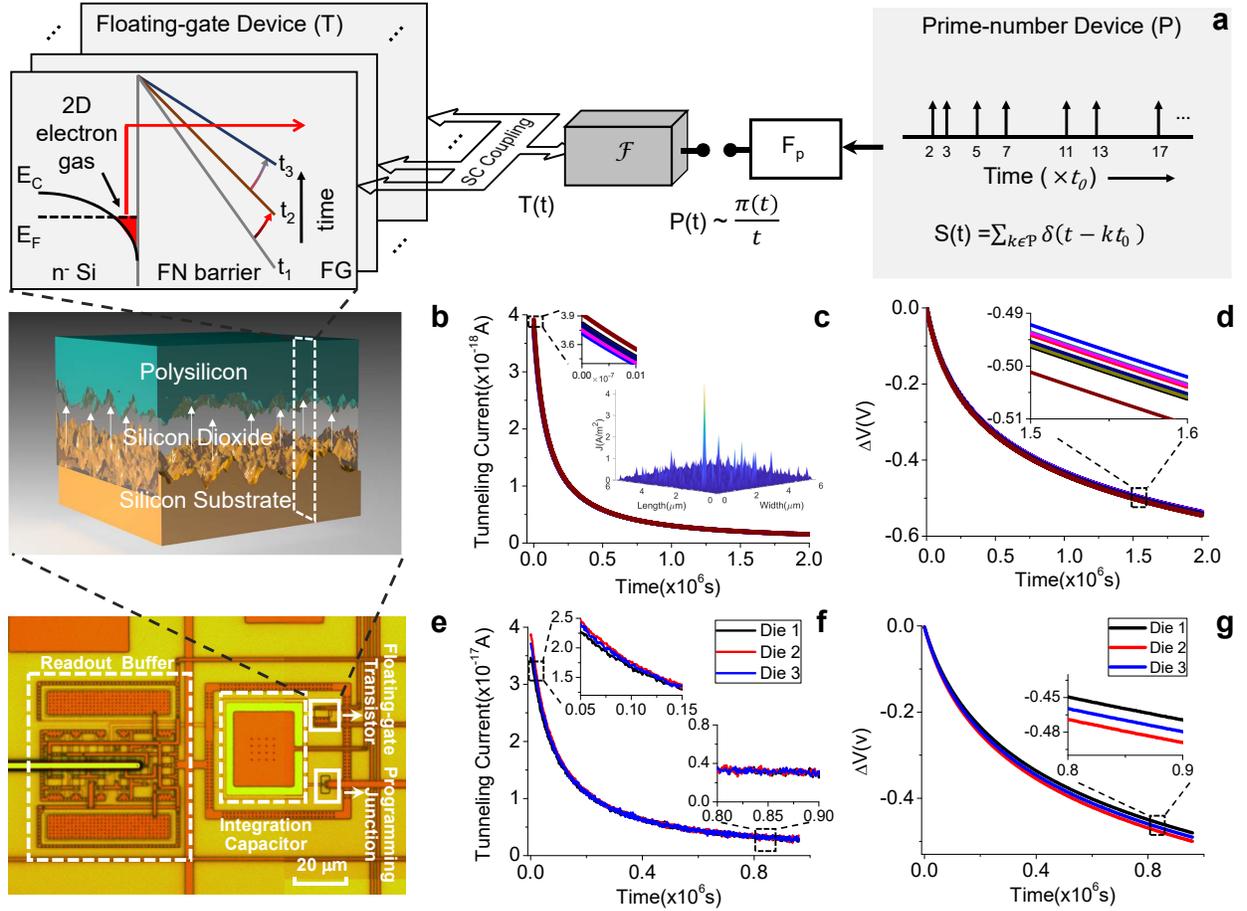}
	\end{minipage}
	\caption{(a) Principle of operation of the FG device (T) where 
	thermally excited electrons tunnel through an array of FN tunneling barriers that are mutually coupled (SC Coupling) to each other. The sequential transport of electrons results in discrete changes in the slope of respective FN barriers at times $t_1$, $t_2$ and $t_3$; The output of the FG device is related to the hypothetical prime number generator (P) through a polynomial transformation $\mathcal{F}$; (b) Practical FG device with electron tunneling through a an array of non-uniform FN oxide-barriers; (c) Simulated tunneling currents for 8 devices with different oxide thickness distribution, showing asymptotic convergence to current distribution shown in inset; (d) Simulated change in FG voltage with respect to time; (e) Micrograph of a fabricated FG device; (f) Measured leakage rates of different FN tunneling devices verifying asymptotic convergence; and (g) Measured change in the FG voltage with respect to time which conforms to the simulated results. }
	\label{fig1}
\end{figure*}

In this work, we have asked an alternative question related to prime-number generation: {\it what type of a dynamical system could produce temporal statistics that matches the asymptotic response as described by prime-number theorems ?} We show in the methods section that at the core of any hypothetical prime-number
generator will be a dynamical system whose physics is governed by Fowler-Nordheim (FN) tunneling of electrons~\cite{lenzlinger1969fowler}. 
We considered a hypothetical system (labeled as P in Fig.~\ref{fig1}(a)) that is driven by a sequence of impulses, with each impulse located at discrete time-instants corresponding to prime numbers. A smoothing filter $F_p$ then processes the impulse sequence to produce the time-varying density of prime numbers $P(t) = \pi(t)/t$, where $\pi(t)$ is the continuous-time variant of the prime counting function that has been extensively studied~\cite{dusart2010estimates,kotnik2008prime}. However, 
practically all forms and bounds concerning the prime density function that have been reported in literature follows an 
asymptotic response $P(t) \approx \mathcal{F}(1/\log(t))$\cite{berry1999riemann,dusart2010estimates,kotnik2008prime}, where $\mathcal{F}$ being a polynomial
function. As shown in the methods section, this type of asymptotics can be naturally generated by first-order dynamical system that is governed by the physics of Fowler-Nordheim (FN) tunneling of electrons followed by a polynomial non-linearity. 

In Fig.~\ref{fig1}(a) we illustrate the dynamics of such a device (labeled as T) where thermally-excited electrons tunnel through the triangular FN tunneling barrier on to a conductive island that is electrically insulated (labeled as FG). 
The tunneling of electrons changes the potential of the island which in turn decreases the slope of the FN tunneling barrier.
At a fundamental level, this device represents a counting process where the sequential tunneling of electrons results in discrete changes in the barrier slope, as illustrated in Fig.~\ref{fig1}(a). 
Therefore, $T(t)$ corresponds to the dynamics of the underlying quantized process that has been smoothed-out due to the process of measurement.
In this work we have fabricated a FN tunneling device and we have compared the statistical properties of $P(t)$ and $T(t)$. 

The FN tunneling device has been implemented using a floating-gate (FG) transistor where the thermally grown gate-oxide (silicon-di-oxide) acts as a FN tunneling barrier that separates the lightly-doped n-type semiconductor and the polysilicon floating-gate (FG). An approximate two-dimensional sheet of electron gas (as shown in Fig.~\ref{fig1}(a)) and a triangular FN tunneling barrier is created by depleting the floating-gate of electrons and creating a large potential difference ($>$7.0 V).
However, due to fabrication artifacts, the oxide and hence the barrier thickness would vary across the entire cross-sectional area of the device. 
As a result, the cumulative electron transport is the combined effect of electrons tunneling through barriers of different thicknesses, as illustrated in Fig.~\ref{fig1}(b). 
The beauty and novelty of the proposed FG device is a self-compensating (SC) physics that asymptotically leads to a configuration where tunneling predominantly occurs through only one spatially isolated barrier (the model for this is summarized in the methods section).
This effect can be clearly seen in Fig.~\ref{fig1} (c) where the Monte-Carlo simulation of the FG device generates the time evolution of the cumulative tunneling currents for different oxide-thickness distributions. 
Initially, the tunneling currents differ from each other (shown in inset of Fig.~\ref{fig1}(c)), however, asymptotically the different time-evolutions converge. 
The spatial distribution of tunneling currents (shown in the inset of Fig.~\ref{fig1}(c) for one specific oxide-thickness distribution) indeed verifies that asymptotically tunneling is dominated by a barrier at one spatial location. 
This effect is also evident in the time-evolution of the floating-gate potential (as shown in Fig.~\ref{fig1}(d)) for different oxide-thickness distributions where
asymptotically, each of the curves converges to a $\approx 1/\log(t)$ response. 
The measured results obtained from a fabricated prototype of an FG device (micrograph shown in Fig.~\ref{fig1}(e)) also verify the Monte-Carlo results. 
Details of the measurement methods and initialization procedures have been provided in the methods section \ref{DeviceProgramming}. 
Note that the tunneling currents for the device asymptotically approach atto-amperes ( $\approx 10^{-18}$ A) which implies an electron tunneling rate around a few electrons per second. 
We fabricated FG devices with different tunneling cross-sectional areas and different floating-gate capacitances and we have verified their respective asymptotic responses at different temperatures. 
The device specifications and the micrographs are shown in the supplementary Fig.~\ref{figs1} which also shows their respective measured responses. 
The results verify that while the transient responses of the devices can be adjusted by changing the device form-factor, the asymptotic response ($\approx 1/\log(t)$) remains invariant across different devices and operating temperature. 
This result also substantiates our previous claim that asymptotically electron transport is dominated by tunneling through a single barrier which results
in a topology that is robust to fabrication mismatch and errors.

\begin{figure*}[ht]
	\centering
	\begin{minipage}[t]{1\textwidth}
		\centering
		\includegraphics[width=1\textwidth]{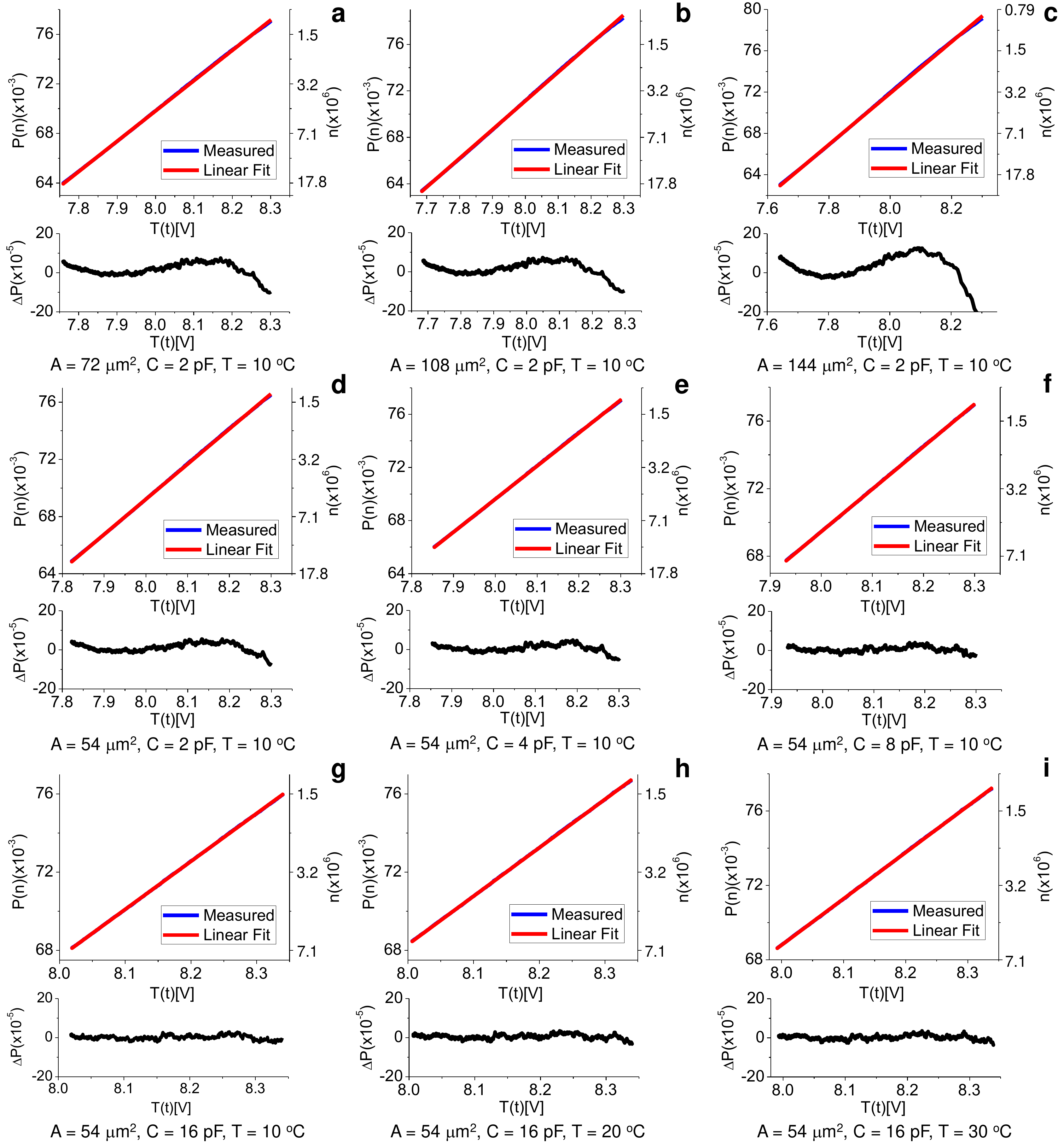}
	\end{minipage}
	\caption{Comparison between the asymptotic prime number density and the output
of different fabricated FG devices operating at different temperatures. (a)-(i) shows linear fit (the main panel) and the corresponding error residues (the bottom panel). For the main panel, the x-axis corresponds to the measured FG device output, the left y-axis corresponds to prime number density P(n) estimated for the total integers (n) plotted on the right y-axis.}
	\label{fig2}
\end{figure*}

We compared the output $T(t)$ of different FG devices with the output $P(t)$ of the hypothetical prime-number device (as shown in Fig.~\ref{fig1}(a)) for prime-number sequences up to 50 million. 
In this case the smoothing filter $F_p$ was chosen to produce the prime-number density function $P(t) = \pi(t)/t$, where $\pi(t)$ denotes the total number of prime numbers up to the integer $n=\lfloor{t/t_0}\rfloor$, $t_0$ is the normalizing time-constant. 
Note that since we are comparing two different processes that occur at different time-scales, we used a temporal alignment and a scaling procedure to compare $T(t)$ and $P(t)$, as illustrated and described in the supplementary Fig.~\ref{figs3}. 
The results summarized in Fig.~\ref{fig2} show near-perfect linear relationship between the asymptotic prime number density and the output of different FG devices. 
However, the respective regression error residues, shown in Fig.~\ref{fig2}, reveal an embedded higher-order polynomial relationship between $T(t)$ and $P(t)$ similar to what has been indicated by the prime-number theorems. 
The magnitudes of the higher-order error residues are determined by the device parameters and operating temperature, thus indicating that there might exist a specific FG device that could show near-perfect agreement between $T(t)$ and $P(t)$.  

\begin{figure*}[ht]
	\centering
	\begin{minipage}[t]{1\textwidth}
		\centering
		\includegraphics[width=1\textwidth]{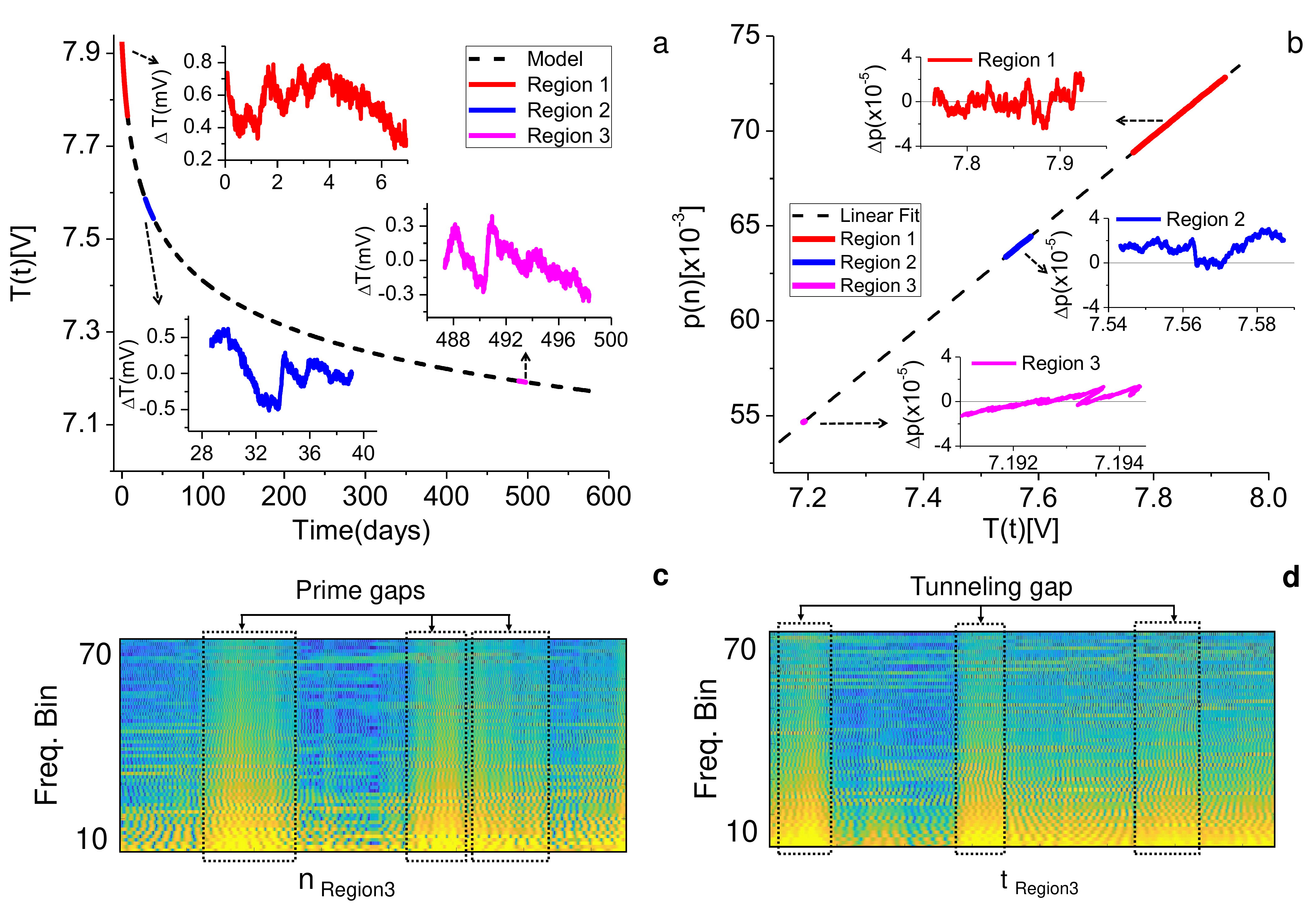}
	\end{minipage}
	\caption{Long-term response of the FG device obtained using a time-stitching technique: (a) Measured response of the FG device
	that follows the $1/\log(t)$ asymptotically and verified for durations greater than 500 days; (b) Linear fit between the device response and the prime number density estimated for 50 million primes. (c) Comparison between the spectrograms of the error residues corresponding to Region 3 clearly showing a non-homogeneous gaps similar to what is observed in the case of prime number sequences.}
	\label{fig3}
\end{figure*}

To further investigate the similarities between the output of the FG device and the density of prime-numbers, we conducted an experiment where we verified if the FG device can continuously operate for durations more than 1 year and yet follow the $1/\log(t)$ response. 
After all, one of the consequences of the prime-number theorems is that there are infinite number of primes. 
If there exists a similarity between the physics of the FG device and prime-number generator, then the FG device should be able to continuously operate over long-durations with matching asymptotic responses. 
We used a time-stitching approach to emulate the operation of the FN device for durations greater than 1 year. 
The result is shown in Fig.~\ref{fig3}(a), where each of the three highlighted regions corresponds to continuous measurement for durations up to 2 weeks, after the FG device has been programmed to different initial voltages. 
The results show that a single $1/log(t)$ time-evolution (shown by dotted line in Fig.~\ref{fig3}(a)) can capture the respective responses in each of the three regions. 
Note that for Region 3, the tunneling rates are in the order of a few electrons per minute and correspond to the device operation up to 1.5 years.
Also note that the magnitude of the residual error shown in Fig.~\ref{fig3}(a) for each of the regions is similar, highlighting that noise-floor is determined by measurement or buffer noise. 
Similar to the previous approach, we compared the measured FN device output to the prime-number device following a time-alignment procedure described in the methods section. 
The results shown in Fig.~\ref{fig3}(b) demonstrate an excellent match between the asymptotic responses of the two processes.  

We compared the spectral statistics of $T(t)$ and $P(t)$ that could reveal a temporal structure in the underlying processes. 
In this regard, Region 3 (in Fig.~\ref{fig3}(a)) is more suitable for analysis since the tunneling rate is in the order of a few electrons per minute. As a result, the discrete-time signatures could be retained after the measurement process of repeated averaging and sampling. 
Details of the data processing are described in the supplementary Fig.~\ref{figs3}. 
For this experiment, we chose the smoothing-filter $F_p$ in Fig.~\ref{fig1}(a) such that it emulates the measurement process of the FG device and also retains some of the temporal statistics of prime number sequences. 
For instance, it is well known that prime-number sequences exhibit large gaps that are void of any prime numbers~\cite{maynard2016chains,maynard2013small}. 
This can be clearly seen in the spectrogram of $P(t)$ in Fig.~\ref{fig3} (c), where the gaps are revealed in time-intervals where only the effect of broad-band measurement noise can be observed. Fascinatingly, similar gaps can be observed in the spectrogram of $T(t)$ in Fig.~\ref{fig3}(d), which implies that there are time-intervals when no electron tunnels through the barrier. This implies that the underlying discrete or quantized process that generates $T(t)$ is not uniform or homogeneous and also exhibits temporal statistics similar to that of a prime number sequence. The non-homogeneity coupled with the $1/\log(t)$
asymptotics of the stochastic process underlying the operation of the proposed FG device motivates future investigation into relationship between
the physics of FN tunneling and the hidden structure of prime-numbers.  

\begin{methods}

\section{Prime number density and FN tunneling}

Consider the hypothetical prime number generator shown in Fig.~\ref{fig1}(a), the output
of the filter $F_p$ is given by
\begin{equation}
\label{PN1}
P(t)= F_p * \sum_{k \in \mathcal{P}} \delta(t-kt_0).
\end{equation}
where $\mathcal{P}$ denotes the set of prime numbers, $t_0$ is a constant time factor
and $*$ denotes a convolution operation. 
Then, as a consequence of the prime number theorem~\cite{dusart2010estimates,kotnik2008prime}
\begin{equation}
\label{PN2}
P(t) = \pi(t)/t = A \left[ \frac{1}{\log(t/t_0)} + \frac{1}{\log^2(t/t_0)} + \frac{2}{\log^3(t/t_0)} + \mathcal{O}(\frac{1}{log^n(t/t_0)}) \right]
\end{equation}
where the parameter $A$ is determined by the gain of the filter $F_p$. 
Denoting $\hat{T}(t)=\frac{1}{\log(t/t_0)}$ the equation~\ref{PN2} can be rewritten as,
\begin{equation}
\label{PN4}
P(t) = A \left[ \hat{T}(t) + \hat{T}^2(t) + 2\hat{T}^3(t) + \mathcal{O}(\hat{T}^n(t)) \right].
\end{equation}
where $\hat{T}(t)$ is a solution of a first-order dynamical system given by
\begin{equation}
\label{PN3}
-\frac{d}{dt} \left[\hat{T}(t) \right]=  \frac{\hat{T}^2(t)}{t_0} \exp\left[\frac{-1}{\hat{T}(t)}\right]
\end{equation}
The right-hand side of equation~\ref{PN3} has a similar
form as the FN tunneling current density $J_{FN}$~\cite{lenzlinger1969fowler} given by:
\begin{equation}
J_{FN}(E) = \alpha E^2 \exp(-\frac{\beta}{E})
\label{FNcurrent}
\end{equation}
where $E$ is the electric-field across a tunneling barrier and $\alpha,\beta$ are material and temperature dependent parameters. 
Assuming an uniform electric-field across the barrier of thickness $t_{ox}$
the first-order dynamical system given by~\ref{PN3} can be implemented using FN current as
\begin{equation}
\label{FNdynamics}
-C\frac{d}{dt} V_{FN}(t) =  \alpha \frac{V_{FN}^2(t)}{t_{ox}^2} \exp\left[\frac{-\beta t_{ox}}{V_{FN}(t)}\right]
\end{equation}
where $C$ denotes the integration capacitance of the floating-gate and $V_{FN}$ is the voltage across the barrier during the process of tunneling. 
The similarity between the two dynamics serves as a motivation to compare $V_{FN}(t)$ and $\hat{T}(t)$ and illustrates a fundamental connection between the density of primes and the proposed FG device. 

\section{Asymptotic tunneling current distribution}

Equation~\ref{FNdynamics} captures the change in floating-gate potential for an uniform oxide or FN barrier thickness.
In practice, the oxide thickness would spatially vary as illustrated in Fig.~\ref{fig1}(b) and can be modeled as a two-dimensional random variable $t_{ox}(x,y)$, with $x$ and $y$ being the two spatial dimensions. 
Then cumulative dynamics in equation~\ref{FNdynamics} can be expressed as the sum of tunneling current over the whole cross-sectional area $-C_T\frac{d}{dt} V_{FN}(t) =  \alpha \sum_{x,y}\frac{V_{FN}^2(t)}{t_{ox}^2(x,y)} \exp\left[\frac{-\beta t_{ox}(x,y)}{V_{FN}(t)}\right]dA$ where $C_T$ is the total floating-gate capacitance and $dA$ is the unit area. 
The first order dynamics guarantees that $dV_{FN}(t)/dt <0$.
It is also highly unlikely that oxide barriers at two different spatial locations will have the same thickness, in which case we can assume that there exists a location $(x_{min},y_{min})$ such that $t_{ox}(x_{min},y_{min}) < t_{ox}(x,y), \forall (x,y) \ne (x_{min},y_{min})$. 
In this case the ratio of the tunneling current
\begin{equation}
\frac{J(t_{ox}(x_{min},y_{min}))}{J(t_{ox}(x,y))} = \frac{t_{ox}^2(x,y)}{t_{ox}^2(x_{min},y_{min})}\exp(\frac{\beta (t_{ox}(x,y) - t_{ox}(x_{min},y_{min}))}{V_{FN}(t)})
\end{equation}
approaches a Dirac-delta function as $V_{FN}(t)$ decreases asymptotically. 
This is what has been verified using Monte-Carlo simulations in Fig.~\ref{fig1}(c) and (d).
For each instance of the simulation, the two-dimensional oxide thickness was generated from a uniform random distribution. 
We used a 100$\times$100 tile array to model the distribution of oxide thickness across a 6$\times$6 $\mu m^2$ cross-sectional area. 
The thickness within each tile is uniform and selected from the uniform distribution with mean value of 13 nm and relative standard deviation of 5$\%$. 
To capture the temporal response, we used equation \ref{FNdynamics} as the dynamics to do numerical integration over time with step size of 1 second. 
The capacitance $C$ is chosen as 2 pF and $\alpha$, $\beta$ are calculated from physics constants and calibrated using measurement data. 

\section{Device implementation}

We fabricated eight different variants of the device on a standard double-poly CMOS process with a $13$nm gate-oxide thickness. Four of these devices have different integration capacitors but identical tunneling junction area, while other four devices have different tunneling junction areas but identical integration capacitors. 
The micro-photograph of the fabricated die is shown in supplementary Fig.~\ref{figs1}(a), and the form factors including the capacitance and tunneling junction areas are all summarized in the table shown in the supplementary Fig.~\ref{figs1}(b). 
Multiple-sections of micrograph were captured using a high resolution microscope and were stitched together as shown in the supplementary Fig.~\ref{figs1}(a). 
The temporal behaviors of all the fabricated devices were measured and also shown in the figure. 
Supplementary Fig.~\ref{figs1}(c) shows the evolution of the tunneling rate with respect to time for each device, where we can observe that the effect of self-compensation (SC) that leads to a converging tunneling response for different devices. 
Supplementary Fig.~\ref{figs1}(d) shows the measured temporal response of all the devices, each of which asymptotically converges to a $1/\log(t)$ response.

\section{Device programming and characterization}
\label{DeviceProgramming}

The device can be initialized to operate in different modes by programming the charge on the floating-gate. The common method for programming FG transistor is by using FN tunneling or by using hot-electron injection~\cite{FGprogramming}.
FN tunneling removes the electrons from FG node by applying a high-voltage ($>$ 15 V for 13nm oxide thickness) across a parasitic nMOS capacitor acting as a programming junction.
Hot-electron injection, however, requires lower voltage ($>$ 4.2 V in 0.5-$\mu$m CMOS process) than tunneling and hence is the primary mechanism for accurate programming of floating-gates. 
The hot-electron programming procedure involves applying a voltage larger than 4.2 V across the source and drain terminals. 
The large electric field near the drain of the pMOS transistor creates impact-ionized hot-electrons whose energy when exceeds the gate-oxide potential barrier ($\approx$3.2 eV) can get injected onto the floating-gate. 
A combination of FN tunneling and hot-electron injection can program the FG voltage to target value.

While the FG voltage ($V_{FG}$) can be easily programmed to the FN tunneling region where electrons can continuously tunnel through the gate oxide, the readout circuits need to be carefully designed to avoid interference with the FG potential. 
Initially, we programmed the FG node voltage to a value around 3V. 
If we want to activate the FN tunneling integration process, we connect the terminal of the integration capacitor ($V_T$) to a fixed voltage such as 6V. 
Since the pMOS transistor is biased at accumulation mode, all the capacitors can be assumed to be constant and FG voltage depends linearly with $V_T$. 
To measure the FG potential, we connect $V_T$ to ground, which pulls the FG voltage below 3V and hence can be interrogated using a standard unity-gain buffer. 
The buffer input is directly connected to the FG node using the same polysilicon layer which avoids the use of metallic vias that might introduce trap states in the surrounding oxide. 

\section{Matching and Alignment}

To time-align the response of the FG device (T) and the prime number generator (P) as shown in Fig.~\ref{fig1}(a), we used the closed-form expression of the FG device model which is obtained from the first-order differential equation~\ref{FNdynamics} reported in~\cite{zhou2017self} and given by: 
\begin{equation}
T(t) = \frac{\gamma_1}{\ln(\gamma_2t + \gamma_3)} + \gamma_4.
\label{device}
\end{equation}
 $\gamma_1$-$\gamma_4$ are model coefficients determined by process parameters, form factors, physics constants and initial conditions. 
After measuring the temporal response of the device, we used the proposed model to fit the data and extract the coefficient $\gamma_{1-4}$. 
A similar process was conducted on the prime number density using the same form as:
\begin{equation}
P(n) = \frac{\lambda_1}{\ln(\lambda_2n + \lambda_3)} + \lambda_4
\label{prime}
\end{equation}
where $\lambda_1$-$\lambda_4$ are model parameters determined by prime number distribution. 
The scalar mapping between the domains (absolute time and prime number generator) was derived using the parameters ($\gamma_{1-4}$, $\lambda_{1-4}$). 
The process is illustrated in supplementary Fig.~\ref{figs3}, where we first obtained the model parameters for the device and the prime number density respectively. 
A linear mapping from absolute time to integer was calculated using those model parameters, which is then used to translate the timer output to prime number density. For alignment, we used the first 50 million prime numbers and  table shown in supplementary Fig.~\ref{figs3} summarizes the corresponding model parameters. 

\section{Time-stitching procedure}

To verify the long-term response of the FG device, we programmed the device in three different regions, as shown in Fig.~\ref{fig3}(a). 
We estimated the device model equation in~\ref{device} using the measured data from Region 1. 
We then extended the model to Region 2 and Region 3 as illustrated by Fig.~\ref{fig3}(a). 
The alignment was achieved by mapping the mid-point of Region 2 and Region 3 to the model value. 
As verified by the results, the model can capture the behavior for a long-term operation equivalent of 1.5 years. 
The model was then used to map the three regions to the prime number density distribution, as shown in Fig.~\ref{fig3}(b) using the mapping method discussed in the previous paragraph.

\section{Spectral analysis}

A spectrogram was used to compare the quasi-stationary fluctuations in the output of the FG device and the output of the
hypothetical prime-number generator. 
We used a 100000 tap finite-impulse-response (FIR) filter $F_p$ to smooth out the impulse train $S(t)$ in Fig.~\ref{fig1}(a). 
This ensures that the output retains some of the high-frequency signatures in the prime-number data without over-smoothing. 
A time-alignment procedure was used to synchronize $T(t)$ and $P(t)$ and a rectangular window (of 100 samples) was used to generate the spectrograms corresponding to the prime number generator and FG device output.

\end{methods}



\bibliographystyle{naturemag}
\bibliography{PrimeNumber}


\begin{addendum}
  \item[Acknowledgements] This research has been supported in part by a research grant from the National Science Foundation (CNS: 1525476). 
 \item[Competing Interests] The authors declare that they have no
competing financial interests.
 \item[Correspondence] Correspondence and requests for materials
should be addressed to Prof. Shantanu Chakrabartty (email: shantanu@wustl.edu).
\end{addendum}

\

\renewcommand{\thefigure}{S\arabic{figure}}

\setcounter{figure}{0}


\begin{figure*}[ht]
	\centering
	\begin{minipage}[t]{1\textwidth}
		\centering
		\includegraphics[width=0.9\textwidth]{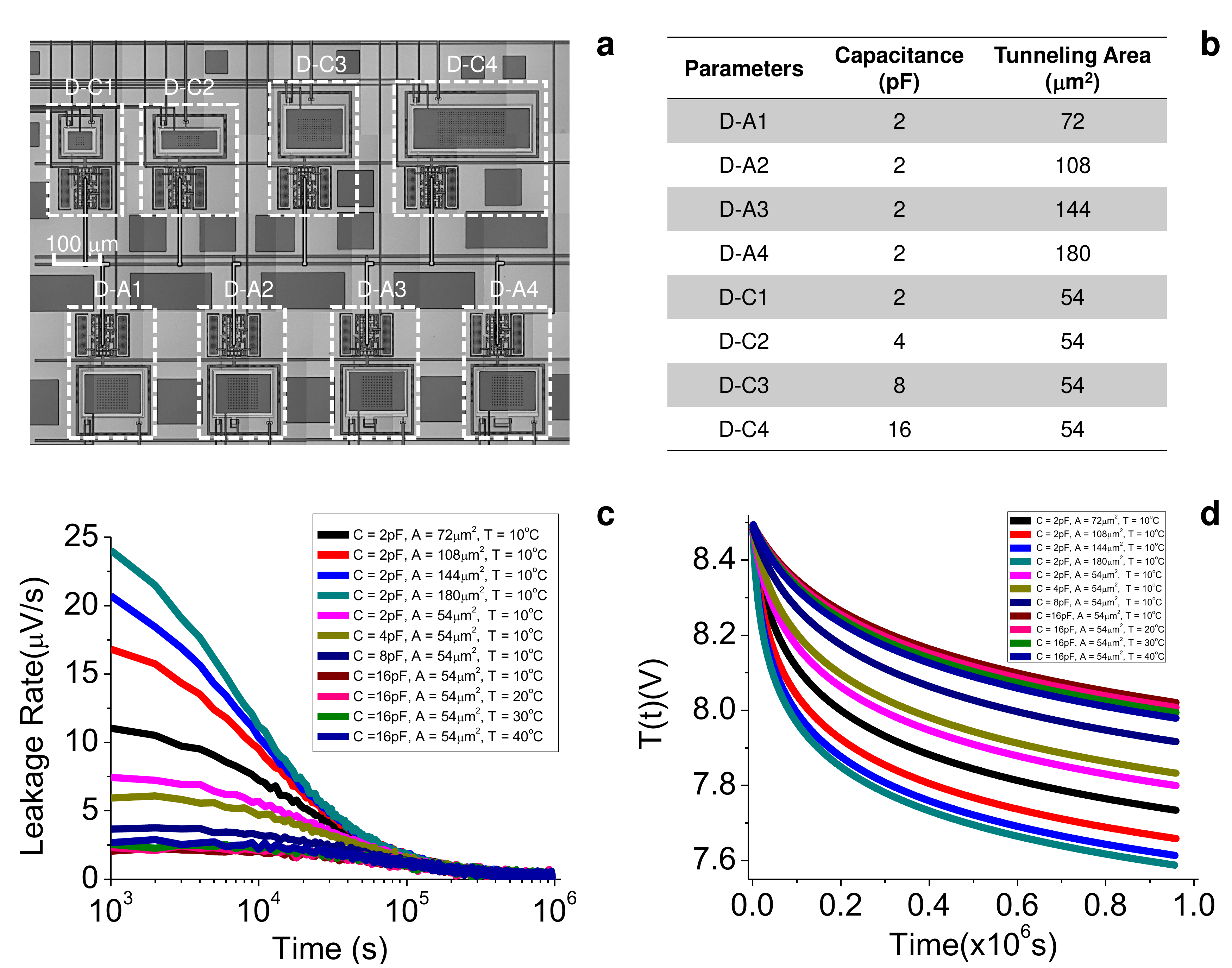}
	\end{minipage}
	\caption{Implementation of the FN tunneling device: (a) die micro-photograph of the fabricated devices with different form factors, (b) summary of the form factors of each fabricated device, (c) measured dependence of tunneling rate on time for each device and (d) measured FG output of the fabricated devices.}
	\label{figs1}
\end{figure*}

\begin{figure*}[ht]
	\centering
	\begin{minipage}[t]{1\textwidth}
		\centering
		\includegraphics[width=0.95\textwidth]{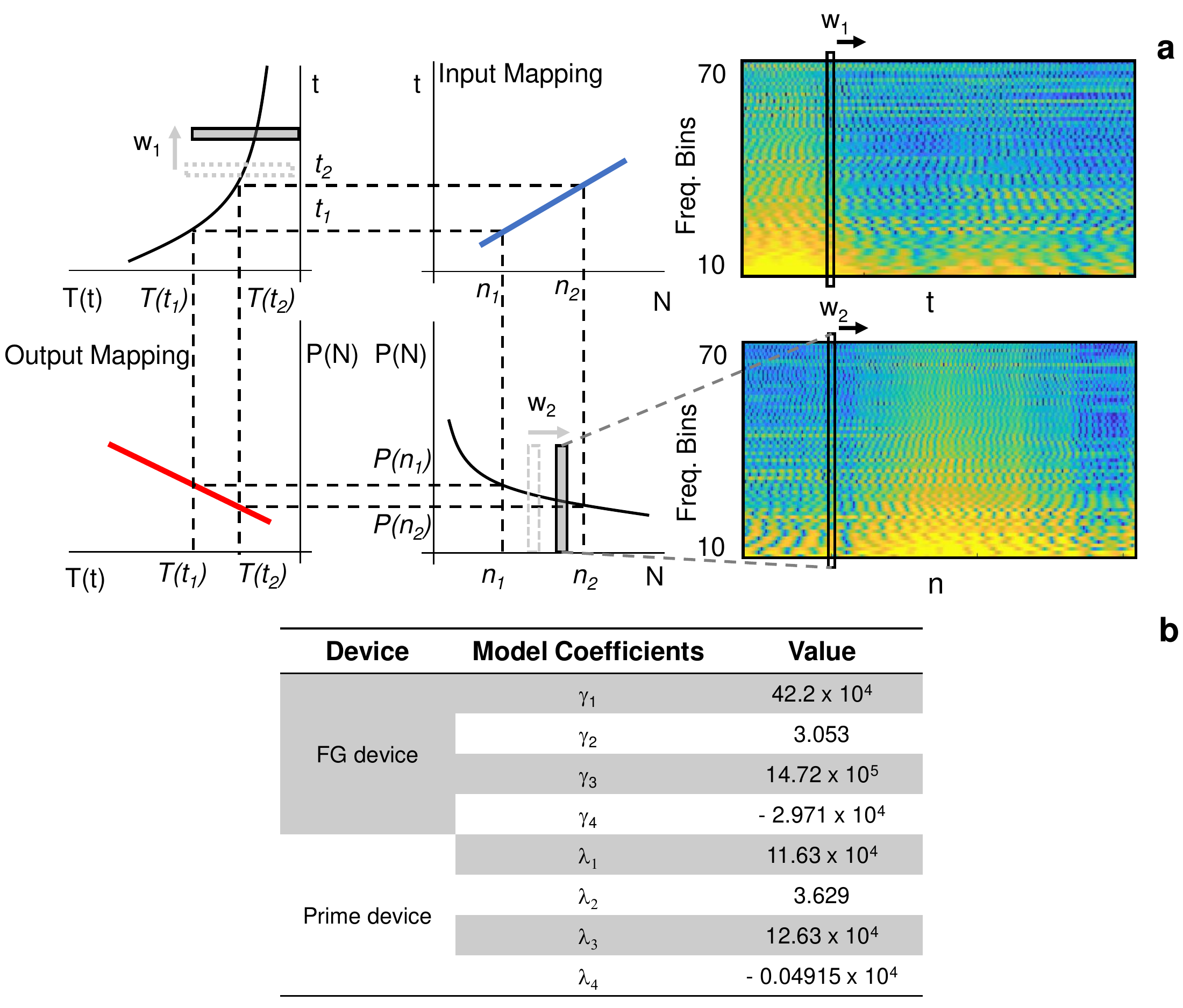}
	\end{minipage}
	\caption{Illustration of the mapping process from the device's temporal response to the prime number density for spectrogram based analysis. (a) Method used for finding the linear mapping between the timer response and the prime number density.  The parameters $\gamma_{1-4}$ and $\lambda_{1-4}$ are used in finding the linear and translation map between $T(t)$ and $P(n)$. The spectrogram is then generated using a similar sized windows chosen from either data. (b) Table summarizing the model parameters for time-alignment and long-term response shown in Fig.~\ref{fig3}.}
	\label{figs3}
\end{figure*}


\end{document}